\documentclass{article}
\usepackage[preprint]{spconf}
\usepackage{amsmath,graphicx}
\usepackage{stfloats, color}
\usepackage{amsmath,graphicx,amssymb,epstopdf}
\usepackage[ruled]{algorithm2e}
\usepackage{multirow}
\usepackage[table,xcdraw]{xcolor}
\usepackage{booktabs}
\usepackage{url}
\usepackage{array}

\newcommand{\Inp}{\mathcal{X}}
\newcommand{\Feat}{\mathcal{Z}}

\newcommand{\Context}{\mathcal{C}}

\title{Learning Invariant Representation and Risk Minimized for \\
	Unsupervised Accent Domain Adaptation}
\name{Chendong Zhao$^{\star \dagger \ddagger}$\thanks{$\ddagger$ Work done during an internship at Ping An Technology. } \qquad Jianzong Wang$^{\dagger * }$\thanks{* Corresponding author: Jianzong Wang, jzwang@188.com.} \qquad Xiaoyang Qu$^{\dagger}$ \qquad Haoqian Wang$^{\star}$ \qquad Jing Xiao$^{\dagger}$}

\address{$^{\dagger}$ Ping An Technology (Shenzhen) Co., Ltd., China \\
	$^{\star}$ The Shenzhen International Graduate School, Tsinghua University, China}

\copyrightnotice{978-1-6654-7189-3/22/\$31.00~\copyright2023 IEEE}
\begin{document}
%
\maketitle
\begin{abstract}
	Unsupervised representation learning for speech audios attained impressive performances for speech recognition tasks, particularly when annotated speech is limited. However, the unsupervised paradigm needs to be carefully designed and little is known about what properties these representations acquire. There is no guarantee that the model learns meaningful representations for valuable information for recognition. Moreover, the adaptation ability of the learned representations to other domains still needs to be estimated. In this work, we explore learning domain-invariant representations via a direct mapping of speech representations to their corresponding high-level linguistic informations. Results prove that the learned latents not only capture the articulatory feature of each phoneme but also enhance the adaptation ability, outperforming the baseline largely on accented benchmarks.
\end{abstract}
\section{Introduction}
\label{sec:intro}

Building inclusive speech recognition systems is a crucial step towards developing technologies that speakers of all accent varieties can use. Therefore, ASR systems must work for everybody independently of the way they speak. Accented speech is indeed a challenge for ASR. Due to the great variability and complexity of accents, it is hard for ASR models to generalize to distinct pronunciations compared to the speeches used for learning~\cite{bell2020adaptation}.
Speech pre-training techniques have emerged as the predominant approach for ASR~\cite{sparse}, and have made speech models much more data efficient: ASR models can be learnt with as little as a few hours of labeled data~\cite{hubert, wav}. For example, wav2vec 2.0~\cite{2} has attained the state-of-the-art performance even available labeled data is few. These gains show that pre-learned features extract linguistic features for improving speech recognition.

\begin{figure}[t]
	\centering
	\includegraphics[width=1.0\linewidth]{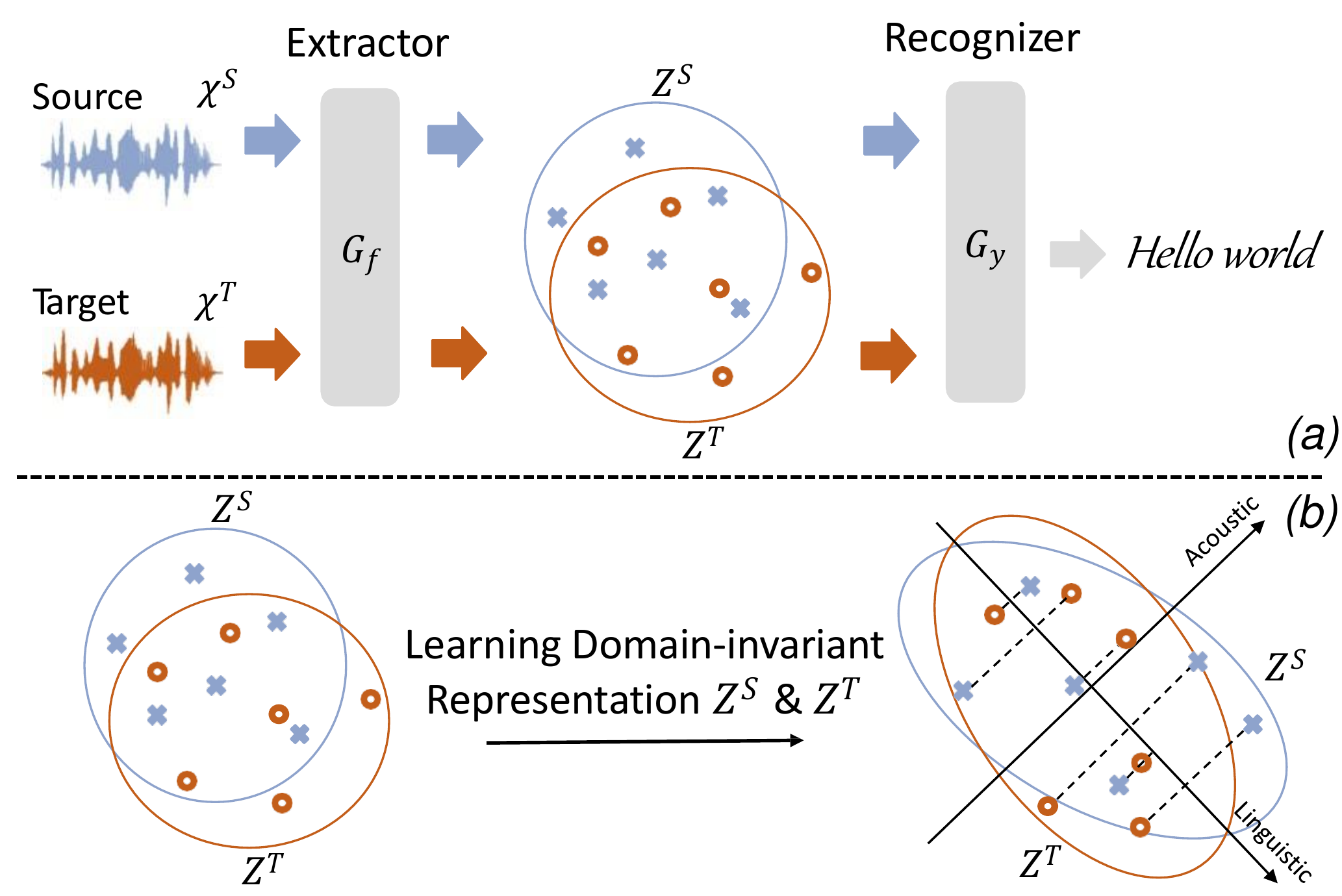}
	\caption{(a) Unsupervised Source-domain learning and Target-domain adaptation both aim to extract powerful representations $Z^S \& Z^T$, which would facilitate subsequent speech-recognization tasks. (b) In this work, we explore learning domain-invariant representations, which aims to eliminate domain-dependent informations (i.e., accent articulations), thus reduces $Z^S \Leftrightarrow Z^T$ distribution discrepancy. Besides, it also contributes to minor risks for better adaptation from source to target.}
	\label{teaser}
\end{figure}

There has been growing interest in unsupervised accent adaptation~\cite{u1,u2,u3}.
Pipeline of unsupervised accent adaptation is shown in Fig.~\ref{teaser} (a). \textbf{First} step is to pretrain on large-scale source speeches $\mathcal{X}^S$. The extractor $G_f$ maps $\mathcal{X}^S \mapsto \mathcal{Z}^S$ via self-learning tasks.
The self-learning objectives, involve future step prediction~\cite{wav}, masked frame represent~\cite{mock,hubert}, and connectionist temporal classification~\cite{CTC}, allow the model to learn intrinsic contextual representations $\mathcal{Z}^S$, which are extremely effective for $G_y$ to recognize texts. \textbf{Second} step continued pretrains $G_f$ on relatively small-scale target (accent) speeches $\mathcal{X}^T$, also by the same self-learning tasks. This step actually is adapting $\mathcal{Z}^S$ to the accent domain, enables the adapted representations $\mathcal{Z}^T$ to also be effective for the accent domain. However, it is not clear how well the adaptation quality $\mathcal{Z}^S \mapsto \mathcal{Z}^T$ to reduce the source-to-target domain discrepancy. Thus, the remaining issue is to find a feature transformation space where $\mathcal{Z}^S \& \mathcal{Z}^T$ distributed as close as possible, i.e., domain-invariant.

However, it is not guaranteed that learning domain-invariant representations is contributive for subsequent recognization tasks.
Meanwhile, it has been observed that discrete latents obtained in an unsupervised learning framework are high-level speech descriptors that correspond to phonemes~\cite{comp,uni,g2p,fedtts}.
Also, vector quantization is widely used in wav2vec-type models~\cite{vqwav,2,3} to create discrete latents. Vq-wav2vec~\cite{vqwav} sets several groups and volumes of codewords to avoid mode collapse, where the context network only uses one mode. Wav2vec 2.0 also follows but introduces with masks. All these works aim to categorize speech encodings by a codebook, thus facilitating subsequent phonemes and words prediction, but relatively little is known about that what linguistic information is encoded by these methods.

Obviously, some form of intermediate representation is needed, able to compensate for accent-differences, and also useful for text recognition. This paper introduces invariant representations, which are learned to model high-dimensional semantic embeddings, eliminating informations about acccented articulations by the way.
So in this work, we propose an \textit{Invariant Representation and Risk Minimized} approach called IRRM, which captures phonetic contrasts while being invariant to properties like the speaker or accent. IRRM is capable of not only learning the articulatory and contextual features of phonemes but also adapting to cross-domain accents well.
We objectively evaluate the extracted representations in the aspect of phoneme recognition and speech segmentation accuracy. Results show that IRRM largely outperforms baseline methods on an accented speech benchmark contributing to the adaptation ability, and as a result, and better mitigates the accuracy discrepancy across domains.
The main contributions are shown as follows:
\begin{itemize}
	\item[$\bullet$] We address the unsupervised accent adaptation in the perspective of Invariant Representation (by explicit phoneme quantization) and risk minimized (by representation distribution alignment), and as a result, better prompts the adaptation performance across domains.
	\item[$\bullet$] Another benefit of IRRM is that quantized frames could be directly clustered to phoneme segments. We propose a cluster correction method after quantization, which can automatically cluster frame-level latents into phoneme-level segments.
\end{itemize}

\section{Related works and Preliminaries}

\subsection{Unsupervised Accent Domain Adaptation}
There are extensive literatures about accent adaptation for speech modeling, and the existing approaches can be largely classified into two mainstreams: accent-independent adaptation and accent-dependent adaptation. The former tends to learn a general speech recognition system which could apply to multiple accents, without the necessary of auxiliary accent-dependent information to guide. For the latter stream, the common approach is to train the model with all kinds of accented speeches under the standard training processs. Moreover, another paradigm is also popular, which follows the multi-task learning manner.
\cite{6} experimented to learn the acoustic model of ASR with an accent recognition classifier jointly, and others introduced to use accent IDs as additional inputs, achieveing an end-to-end ASR system for both recognizing contents and accents.
Another branch, different from multi-task manner, is utilizing the gradient reversal calculation under the adversarial training paradigm. It trains the acoustic model to learn independent informations over accents.
Overall, The major motivation of accent-dependent methods \cite{9} aim to explore the accent-correlative features for auxiliary informations, including i-vector and x-vector, enabling the system generalizable to various accents, or to fine-tune a general recognizer on each accent domain.

\subsection{Unsupervised Speech Representation Learning}

Unsupervised representation learning paradigm is introduced in the speech field to learn representations, which are contributive to downstream tasks, including phoneme recognition and semantic discrimination~\cite{lai2021semi}. It levelages the advantage of amounts of unlabeled speech from pre-training tasks which explore general characteristics. Pre-training tasks could be largely classified into two branches: speech reconstruction and prediction. Speech reconstruction is usually employed in auto-encoder models~\cite{hsu2017learning}, in which speech firstly encoded into latent vectors, and decoded to the original form. Also, many restrictions were introduced into the latent vector, like temporal consensus~\cite{khurana2020convolutional}, numerical discreteness~\cite{ondel2016variational}, and retaining distinction~\cite{hsu2017unsupervised}.
Speech prediction branch trains to predict informations of maksed parts relying on their contexts. The informations to predict consist of spectrograms, group indexes~\cite{ravanelli2020multi}, and contrastive labels of whether the calculated is the masked frame.

\section{Proposed Methodology}
\label{pro}

\begin{figure}[t]
	\centering
	\includegraphics[width=1.0\linewidth]{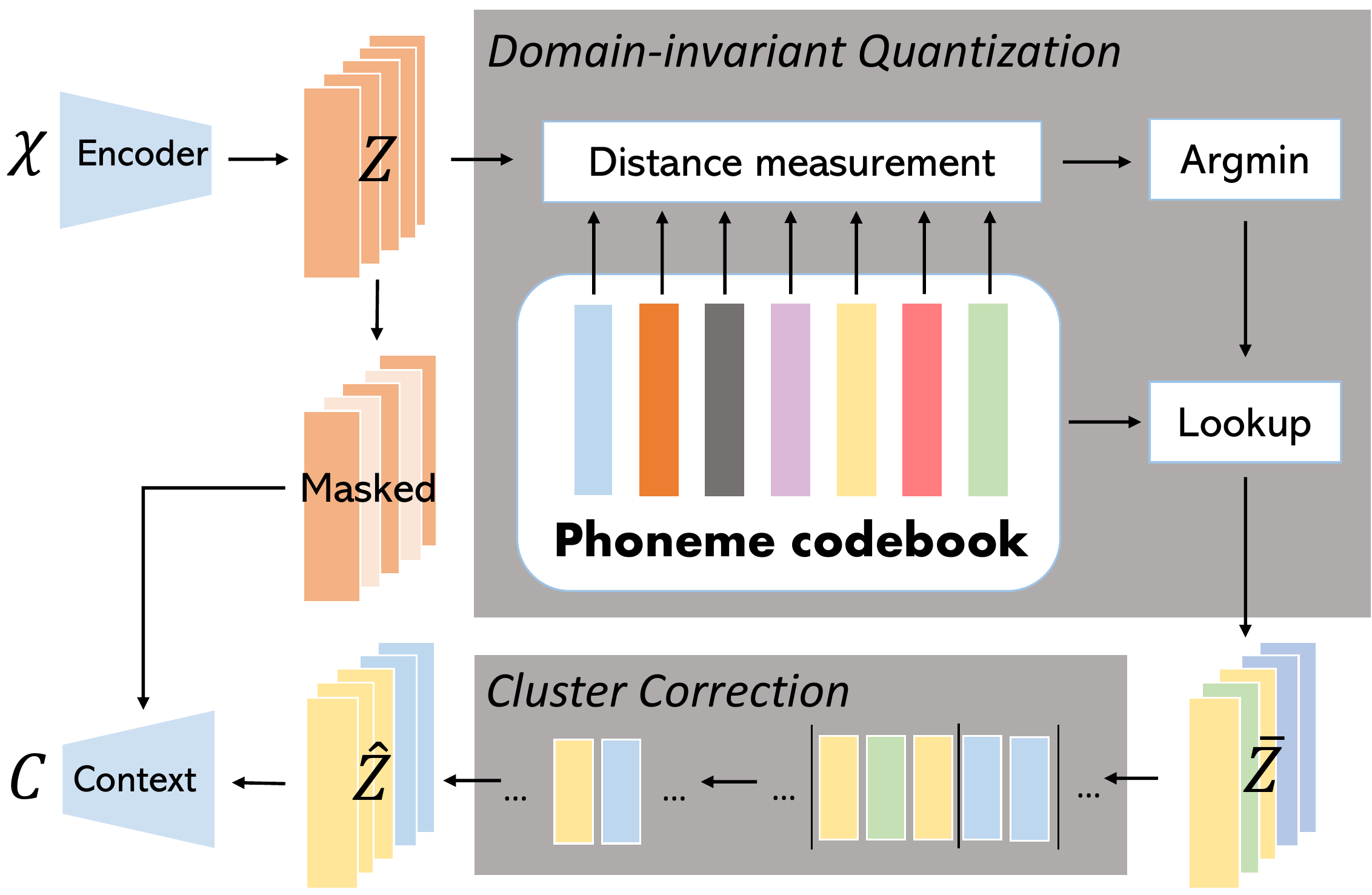}
	\caption{For unsupervised speech representation learning, IRRM uses an encoder and a quantizer to extract domain-invariant features $\bar{z}_{t}$. Then we correct $\bar{z}_{t}$ into $\hat{z}_{t}$ by cluster-context information, finally feed $\hat{z}_{t}$ to the context network.}
	\label{fig:VQ}
\end{figure}

We follow the wav2vec 2.0~\cite{2} pre-training pipeline and consider it as our baseline.
Wav2vec 2.0 contains a convolutional extrator $f: \Inp \mapsto \Feat$ that first encodes the input waveform (of $T$ time steps, $t \in T$) to latent speech representations. These representations are then fed into a Transformer model $g: \Feat \mapsto \Context$ to calculate the context features $\Context$. Specifically, a portion of $z_t$ are masked and then inputted into the context model.
The other branch $\Feat$ are quantized to $\bar{\Feat}$ via the quantization component $\Feat \mapsto \bar{\Feat}$. The quantization component utilizes the Gumbel softmax calculation to assign codewords from $G=2$ codebooks, each contains $V=320$ entries, and the assigned codewords are concatenated to have $\bar{z}$. The overall training aims to predict the true quantized vector $\bar{z_t}$ among $c_t$, which are masked distractors from other time steps. As shown in Fig~\ref{fig:VQ}, this work proposes:

\noindent(1)
To acquire invariant representations, we define the codebook to correspond with linguist information, so to eliminate accent-dependent articulatory differences.
IRRM codebook only applies 40 codewords, 39 assigning to phonemes according to CMU phoneme set~\cite{CMU} and one denoting silence.

\noindent(2) After quantization, IRRM corrects the quantized $\bar{z}_t$ with the help of cluster-context, then inputs the corrected $\hat{z}_t$ into the following context network. We assign articulatory-similar frames to the same codeword, then group consecutive latents into segments, transforming the frame-synchronized representations into phoneme-synchronized sequences.

\subsection{Learning Invariant Representation}
\label{Phonetic quantization}
The goal of IRRM of learning phoneme units is to capture articulatory contrasts while being invariant to properties like the speaker or accent, i.e., domain-invariant.

\subsubsection{Invariant Representation Setup}

The codewords remain meaningless since the input signal does not force them to denote phonemes.
So in this subsection, we will theoritically show that how each codeword of the codebook could be assigned to a phoneme, with the help of a few minutes of time-aligned annotated phonemes $\left(X_{\text {pair }}, Y_{\text {pair }}\right)$.
Firstly, we adjust the size $V$ of the codebook $E=\left\{e_{1}, e_{2}, \ldots, \notag\right. \left. e_{V}\right\}$ cover all phonemes.
Each codeword $e_v$ in $E$ is mapped to denote a specific phoneme $v$.
As the encoded $z_t$ has continuous values, the quantifying probability of it assigning to a codeword $e_v$ would be calculated as:

\begin{equation}
	P\left(v \mid z_{t}\right)=\frac{\exp \left(-\left\|z_{t}-e_{v}\right\|_{2}\right)}{\sum_{k \in V} \exp \left(-\left\|z_{t}-e_{k}\right\|_{2}\right)}.
\end{equation}

\noindent Also, the probability of $z_t$ belonging to one of all phonemes $\tilde{Y}=\left(v_{1}, v_{2}, \ldots, \notag\right.\left. v_{T}\right)$ could be measured as:

\begin{equation}
	P(\tilde{Y} \mid Z) \approx \prod_{t=1}^{T} P\left(v_{t} \mid z_{t}\right).
\end{equation}

\noindent but the product of the above approximations for $T$ frames doesn't match the length of the phoneme-synchronized sequence $Y_{pair}$, which only has $S$ phonemes. Because each phoneme may correspond to several repeated codewords $e_v$. We solve this problem by considering the connectionist temporal classification (CTC)~\cite{CTC} task, so from CTC we get:

\begin{equation}
	P\left(Y_{\text {pair }} \mid Z\right)=\sum_{\tilde{Y} \in Y^{\prime}} P(\tilde{Y} \mid Z),
\end{equation}
where $Y^{\prime}$ involves all potential $\tilde{Y}$ which reduces to $Y_{pair}$, by inserting blank symbols to $Y_{pair}$ until it reaches length $T$.
The mapping operation not only brings interpretability to the codebook but also prevents some codewords from collapsing while some other codewords learn nothing.

\subsubsection{Invariant Representation Quantization}
After mapping the codewords to their corresponding phonemes, we perform the quantization learning process for the pre-training speeches.
For each time step $t$, a k-means (KM) codebook is conducted by substituting the output of the encoder vector $z_t$ with its closest codeword (in Euclidean Distance) within the codebook.
Formally, we quantize each $z_{t} \in Z$ to become one entry from a trainable vector book $E = \left(e_{1}, e_{2}, \ldots, e_{v}\right)$, where $e_v$ being an entry in the codebook $E$, with size $V$. 
Since selecting the closest entry makes the whole quantization module become non-differentiable, during during back propagation, the gradient passed to the encoder would be estimated by a straight-through estimator:
\begin{equation}
	\bar{z}_{t}=z_{t}+e_{v}-\operatorname{sg}\left(z_{t}\right), \text { where } v=\underset{k}{\arg \min }\left\|z_{t}-e_{k}\right\|_{2},
\end{equation}
where $\operatorname{sg}(\cdot)$ being a stop-gradient calculation which regards its input as invariant during back propagation.

In addition, there's another version, the Gumbel-Softmax (GS)~\cite{gumble,ququ}. It hard-chooses one entry $e_v$ using a linear projection $l \in \mathbb{R}^{V}$. Probability for selecting j-th codeword is:
\begin{equation}
	p_{j}=\frac{\exp \left(l_{j}+r_{j}\right) / \tau}{\sum_{v=1}^{V} \exp \left(l_{v}+r_{v}\right) / \tau},
\end{equation}
where $r=-\log (-\log (u))$ and $u \sim U(0,1)$, $\tau$ adjusts the Gumbel-Softmax temperature.
During model's forward calculation, a hard index $\operatorname{argmax}_{j} p_{j}$ is chosen,
and during the backward propagation, the real gradients are derived from the softmax, making the code selection completely differentiable and all codewords updated.

The quantization is essentially an extreme form of sparseness in the latent space.
In wav2vec 2.0-way of pre-training, the codewords would robustly match the prominent articulatory features of each phoneme.

\subsection{Cluster Correction}
\label{cl}
After phoneme quantization, many frame latents $z_t$ adjacent in time with similar articulation may be quantized to the same codeword $e_{v}$.
However, it is inevitable that some latents would be incorrectly quantized, wrongly assigned with other irrelevant phonemes.
To alleviate and reduce the chances of these quantization distortions, we propose to detect and correct them by cluster-context informations.

We leverage recent advances in anomaly detection~\cite{anoma}, using a k-nearest neighbor (kNN) classifier to identify anomalous (low-likelihood) patterns. This method achieves powerful performances when dealing with semantic high-dimensinal representations, which are easily feasible in this task.
Given quantizied latents $\bar{\Feat}=\left(\bar{z}_{1}, \bar{z}_{2}, \ldots, \bar{z}_{T}\right)$, for each time step we sample a series, centered at $t$ and related a window length of $win$. We refer as $\widetilde{z}_t = [\bar{z}_{t - \frac{win}{2}}, \ldots, \bar{z}_t, \ldots, \bar{z}_{t + \frac{win}{2}}]$. This segment window is concatenated by all other representations calculated from the segment and lastly flattening, formed in a $win * d_e$ dimensional vector. $d_e$ is the dimension of $\bar{z}_t$ ($768$ for wav2vec 2.0).
The anomaly detection is measured via the kNN criterion, which calculates distances of $\widetilde{z}_{t}$ and others within same sample. This can be formal as:
\begin{equation}
	s_t = \sum_{\widetilde{z} \in {\cal{F}}_{kNN}} \|\widetilde{z} - \widetilde{z}_{t}\|^2_2,
\end{equation}
where $s_t$ being the criterion value and ${\cal{F}}_{kNN}$ denoting the training set of $k$ representations, resulting in the distance measurement of $\widetilde{z}_{t}$ to be minimal.
Lastly, a series of anomaly values is calculated to be $\widetilde{\mathbf{s}} = \{s_1,s_2 \ldots s_T\}$.
Phoneme boundaries thus are chosed from these candidates.
Specifically, the motivation is from that the most anomalous points are assumed to be correlating with boundaries. Therefore, we employ a peak detector\footnote{ We use the standard peak detection function by the \textsc{Scipy} package.} to detect the most peak anomaly time point. Also, we constrain that adjacent peak times should be no nearer than 60 ms (since few phonemes having shorter duration than 60 ms).
In the final stage, each segment within boundaries corresponds to a phoneme unit.
Each latents $\bar{z_t}$ within the same phoneme unit is replaced by the majority within the unit, transforming $\bar{\Feat}$ to $\hat{\Feat}$, such that wrongly quantized latents would be corrected with the help of cluster-context informations.

Another benefit is that this approach achieves unsupervised phoneme segmentation.
Since the unsupervised segmentation is a pivotal prerequisite in many speech processing tasks, and also enables language acquisition without annotations, this would encourage us to explore lower or zero-resource accent speeches in the future work.

\subsection{Risk Minimized Accent Adaptation}

Although the extracted invariant representation distributions are assumed to be identical across domains, we further explore to enhance the consistency during the accent adaptation phase.
Specifically, we propose to regularize the phoneme codebook updating under risk minimization constraints.
We firstly follow the baseline masked prediction loss. After we get the quantized $\hat{\Feat}$, the context network then combines sequences of $\hat{z_t}$ and contextual representations $\Context$. The context module is learned to identify the true $\hat{z_t}$ among a series of distractors from other masked time steps:
\begin{equation}
	\mathcal{L}^{\text {wav2vec 2.0}}=-\log \frac{\exp \left( sim \left(c_{t}, \hat{z_t}\right)\right) / \kappa}{\sum_{\hat{z} \in \Omega} \exp \left( sim \left(c_{t}, \hat{z}\right)\right) / \kappa},
\end{equation}
where $\Omega$ is the set of $\hat{z}_{t}$, sampled from $N$ negative time-steps, $\kappa$ controls the temperature and $sim$ utilizes the cosine calculation $sim(m, n)=\frac{m^{T} n}{\|m\||\| n|}$.
As the codewords consistency between the source domain and target domain enhance the adaptation ability of the quantization policy, we constrain the codebook adapting, so there would be an additional loss term incorporated for adaptation Risk Minimization:
\begin{equation}
	\mathcal{L}^{rm}=\left\|s g\left(z_t\right)-\hat{z_t}\right\|_{2}+\beta\left\|z_t-s g\left(\hat{z_t}\right)\right\|_{2},
\end{equation} 
where first term moves codewords closer to outputs of the encoder, and the second term enforces encodings to commit to codewords. $\beta$ is an hyperparameter. We minimize the primary masked prediction loss together with $\mathcal{L}^{rm}$:
\begin{equation}
	\mathcal{L}^{total}=\mathcal{L}^{\text{wav2vec}}+\gamma \mathcal{L}^{rm},
\end{equation}
where $\gamma$ controls the penalty of codebook regularization. The regularization loss aims to enhance the consistency between encodings and codewords, guiding the updates stably when adapting to accented data distribution.
When dealing with accented speeches of specific pitches, prosodies or temporal environment noises, the learnable codewords would adapt to these new features, so to model the accented phonemes.

\section{Implementation Details}
\label{implentation}

\subsection{Dataset}

\noindent \textit{1) Source Domain:} We use Librispeech 960h~\cite{libi} to pre-train IRRM and wav2vec 2.0 base model, which is considered as the source domain (US English). We evaluate models in-domain performance, where we apply the standard evaluation protocol of TIMIT dataset~\cite{timit} and consider 39 phonemes.
TIMIT consists of 5 hours of speech audios, which are annotated by time-aligned phonemes. We split it with the standard train/dev/test portion, and use a partial split for mapping data.

\noindent \textit{2) Accent Domain:} For cross-domain accents, we use the CommonVoice~\cite{CV}, an accented open source collected by Mozilla.
We gathered German (DE), British (UK), Indian (IN), and Australian (AU) raw accented speeches.
For each accent, we conduct the same finetune strategy, via same adaptation data (25 speakers, 2.5 hour) and same adapting steps (20k). Also, we collected almost the same amounts of test data (1h and random unseen speakers).

\subsection{Model Configurations}

We adapt the fairseq~\cite{seq} implementation of wav2vec 2.0.
The encoder architecture is the same as the baseline method, except for the last layer's ouput to match the quantizer.
$V = 40$ is set as mentioned, where each codeword is a randomly initialized vector of 768-d.
The gumbel temperature $\tau$ is linearly annealed from 2 to 0.5 following the baseline setting.
We ablated and chose the following hyper-parameters in the cluster correction: The number of neighbors is k = 20 and $win$ = 10.
We determine the loss coefficients by grid search. $\beta$ and $\gamma$ are 2 and 0.5, respectively.
We use the wav2letter++ toolkit~\cite{w2l} as our acoustic models.
In this work, we mainly focus on the phonemes modelling ability (PER (\%)) and omit further word prediction tasks which are affected by language models.
Phoneme error rate is calculated by $PER = \frac{D + S + I}{T}$. T is the whole number of phonemes, and D for wrong deletions, S for wrong substitutions, I for wrong insertions.


\section{Results}
\label{results}

\begin{table*}[t]
	\caption{Average PER (\%) by Standard Error (SE) statistics on CommonVoice Testset in the cross-domain results. Adp. and RM. denote whether adaptation and the risk minimized loss, respectively. Avg. is computed on all test data.}
	\centering
	\label{acc}	
	\resizebox{0.86\linewidth}{24mm}{
		\begin{tabular}{c|cc|cccc|c}
			\toprule[1.5pt]
			&  \text{Adp.} & \text{RM.} & \text{UK} & \text{DE} & \text{IN} & \text{AU} & \text{~~~Avg. PER (\%) } \\ \midrule[1pt]
			\multicolumn{1}{c|}{\multirow{2}{*}{\text{ wav2vec 2.0~\cite{2}, KM}}} & -- & & 34.8 $\pm$ 1.46 & 34.5 $\pm$ 1.06 & 39.6 $\pm$ 1.00 & 40.1 $\pm$ 0.82 & 37.3 $\pm$ 1.08     \\
			\multicolumn{1}{c|}{} & \checkmark & & 32.8 $\pm$ 0.74 & 33.4 $\pm$ 0.72 & 37.8 $\pm$ 0.78 & 36.2 $\pm$ 0.82 & 35.1 $\pm$ 0.97\\
			\multicolumn{1}{c|}{\multirow{2}{*}{wav2vec 2.0~\cite{2}, GS}} & -- &  &32.4 $\pm$ 0.83 & 31.1 $\pm$ 0.68 & 31.7 $\pm$ 1.06 & 41.3 $\pm$ 0.64 & 33.5 $\pm$ 0.67   \\
			\multicolumn{1}{c|}{} &  \checkmark &  &27.9 $\pm$ 1.25 & 35.6 $\pm$ 1.19 & 28.1 $\pm$ 1.22 & 28.8 $\pm$ 1.01 & 30.4 $\pm$ 1.18  \\
			\midrule
			\multicolumn{1}{c|}{\multirow{3}{*}{\text{ IRRM, KM}}} & -- & -- & 40.3 $\pm$ 0.94 & 39.1 $\pm$ 1.01 & 44.6 $\pm$ 1.03 & 42.2 $\pm$ 0.98 & 41.6 $\pm$ 1.04 \\
			\multicolumn{1}{c|}{} & \checkmark &-- & 32.4 $\pm$ 0.71 & 30.9 $\pm$ 0.62 & 29.5 $\pm$ 0.79 & 29.6 $\pm$ 0.77 & 30.3 $\pm$ 0.54 \\
			\multicolumn{1}{c|}{} & \checkmark & \checkmark & 30.1 $\pm$ 1.41 & 28.5 $\pm$ 1.32 & \textbf{27.6 $\pm$ 1.35} & 29.2 $\pm$ 1.36 & 29.4 $\pm$ 1.48   \\
			\multicolumn{1}{c|}{\multirow{3}{*}{\text{ IRRM, GS}}} & -- &-- & 41.1 $\pm$ 0.77 & 38.6 $\pm$ 0.51 & 43.7 $\pm$ 0.66 & 41.7 $\pm$ 0.98 & 41.3 $\pm$ 0.72   \\
			\multicolumn{1}{c|}{}  & \checkmark & -- & 30.3 $\pm$ 0.56 & 33.3 $\pm$ 0.62 & 33.1 $\pm$ 0.67 & 34.5 $\pm$ 0.76 & 32.8 $\pm$ 0.69  \\
			\multicolumn{1}{c|}{} & \checkmark & \checkmark & \textbf{26.5 $\pm$ 0.70} & \textbf{27.4 $\pm$ 0.88} & 28.9 $\pm$ 0.69 & \textbf{26.7 $\pm$ 0.71} & \textbf{27.4 $\pm$ 0.82} \\ 
			\bottomrule[1.5pt]
	\end{tabular}}
\end{table*}

\subsection{Overall Evaluation}

\subsubsection{Cross-domain Evaluation}

For the adaptation ability analysis, we evaluate our method on accented English speech.
All comparing methods are pre-trained and fine-tuned on the same amounts of data and iterations.
Results are shown in Table~\ref{acc}.
Despite the small amount of adaptation data, the overall PER improvement resulting from the combination of our two codebook adaptation techniques reaches almost 10\% relative for accented speech.
Our method shows great adaptation ability, achieves average -14.0 \% after adaptation for gumbel-based approach and -12.2 \% for k-means. IRRM exceeds baselines by -3.1 \% and -6.2 \% for two approaches.
The GS approach shows better generalization than KM, contributing to the end-to-end back-propagation.
Moreover, the risk minimized loss indeed facilitates the codebook's adapting and improves performances further.
Our smaller but effective codebook avoids overfitting to in-domain and enhances adaptation ability to accents.

\subsubsection{In-domain Evaluation}

To evaluate the impact of different amounts of mapping data on the modeling ability, we conduct ablation experiments by different codebook settings. In the IRRM setting, we omit the cluster correction.
All reported here use GS approach.
Results in Tabel~\ref{mp} obviously show the superiority that the cluster correction brings. The gap between with and w/o cluster correction becomes larger when mapping data decreses, which shows that IRRM can significantly improve modeling ability when transcribed data is extremely rare.
On the other hand, the accuracy steadily increases when more mapping data was used for mapping.
Although there are just few minutes of mapping data, we could observe a rapid improvement for the modeling ability.
As validated in~\cite{2}, the capacity matters a lot for the codebook’s modeling ability.
Considering that baseline (2 group, 320 codewords, 1.664k compression bitrate) is extremely larger than ours (40 codewords, 0.53k compression bitrate), we believe a 3-4 \% difference is acceptable.
This proves that IRRM can represent speech contents with much less parameters, but still efficient enough.

\begin{table}[h]
	\caption{PER (\%) varying amounts of mapping data.}
	
	\label{mp}
	\centering		
	\resizebox{\linewidth}{8mm}{
		\begin{tabular}{lcccc}
			\toprule & \text { 37.0 min } & \text { 20.5 min } & \text { 10.0 min } & \text { 3.2 min } \\
			\midrule \text { wav2vec 2.0 2*320 codewords } & 21.20 & 27.13 & 38.49 & 46.54 \\
			\midrule \text { IRRM } & 27.26 & 34.03 & 48.81 & 59.69 \\
			\text { IRRM + Cluster Correction } & 23.64 & 30.09 & 42.25 & 49.32 \\
			\bottomrule
	\end{tabular}}
\end{table}

\subsection{Invariant Representation Visualization}
\begin{figure}[t]
	\centering
	\includegraphics[width=1.0\linewidth]{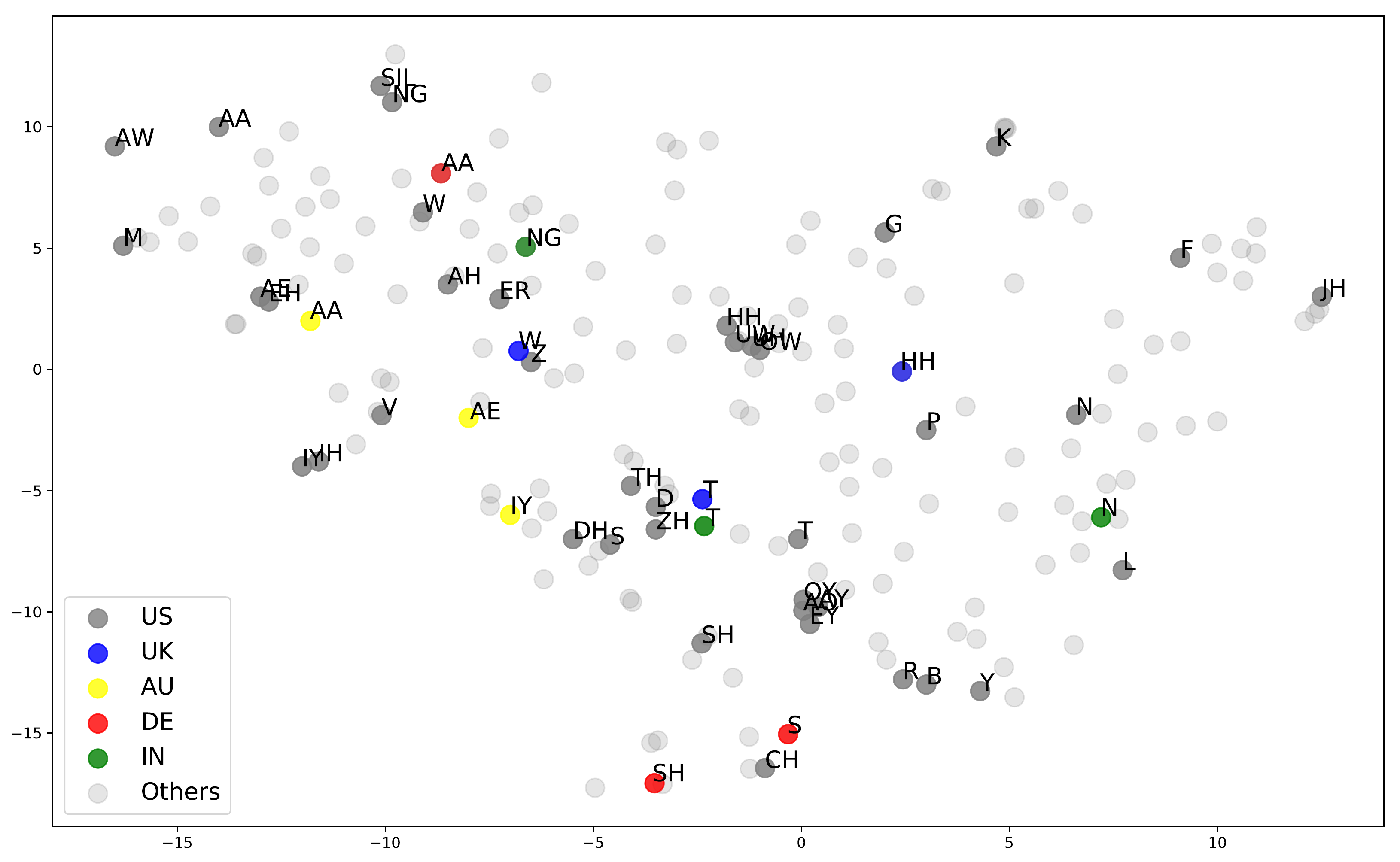}
	\caption{t-SNE ($\rho$=40, $\nu$=10, $\epsilon$=1600) visualization for phoneme codewords. US English pretrained are in deep gray and accent-adapted ones are in colors.}
	\label{fig:ph}
\end{figure}

In Figure~\ref{fig:ph}, we visualize the 39 phonemes' codewords and the silence code by t-SNE~\cite{sne}.
For each accent, we select 3 biggest changed codewords (in euclidean distances) and visualize them all.
From an overview, almost all vowels group on the left-top of the figure, and strong-stress pronounced phonemes stay at bottom. We could see some collapses for some phonemes, like "AY" and "EY", "AE" and "EH", both sounds similar by human ears.
After the adaptation, there's a substantial change for some codewords. For instance, the India and British accents move "T" toward "D" from the original position, which is a notable pronouncing feature for both accents~\cite{IN}; Australia accent treats vowels differently, australians pronounce vowels foward in the mouth (sometimes curled back, in the so-called retroflex position)~\cite{AU}, this character is also appeared in the figure as an integral movement for vowels.
These phenomenons greatly prove that our codewords is capable of matching the spoken-character of phonemes.

\subsection{Unsupervised Phoneme Segmentation}

In contrast to the baseline, whose codewords are meaningless, IRRM could automatically cluster adjacent frames into phoneme segments. Also, the segmentation accuracy would be a validate estimation of our phoneme representation.  
We measure precision, recall, F-score and over-segmentation R-value metric of boundaries with a tolerance of 10 ms in Table 3.
For comparing method, a greedy approach (denoted as ‘Greedy N-seg.’), where the closest adjacent codes are merged until a set number of segments are reached.
%
For the IRRM row, repeated codes after quantized are simply collapsed.
After all hyperparameters tried, we found \textit{Window length} of $10$ frames collaborating with \textit{Number of nearest neighbors} of $5$ achieves the best result.

\begin{figure}[h]
	\centering
	\includegraphics[width=1.0\linewidth]{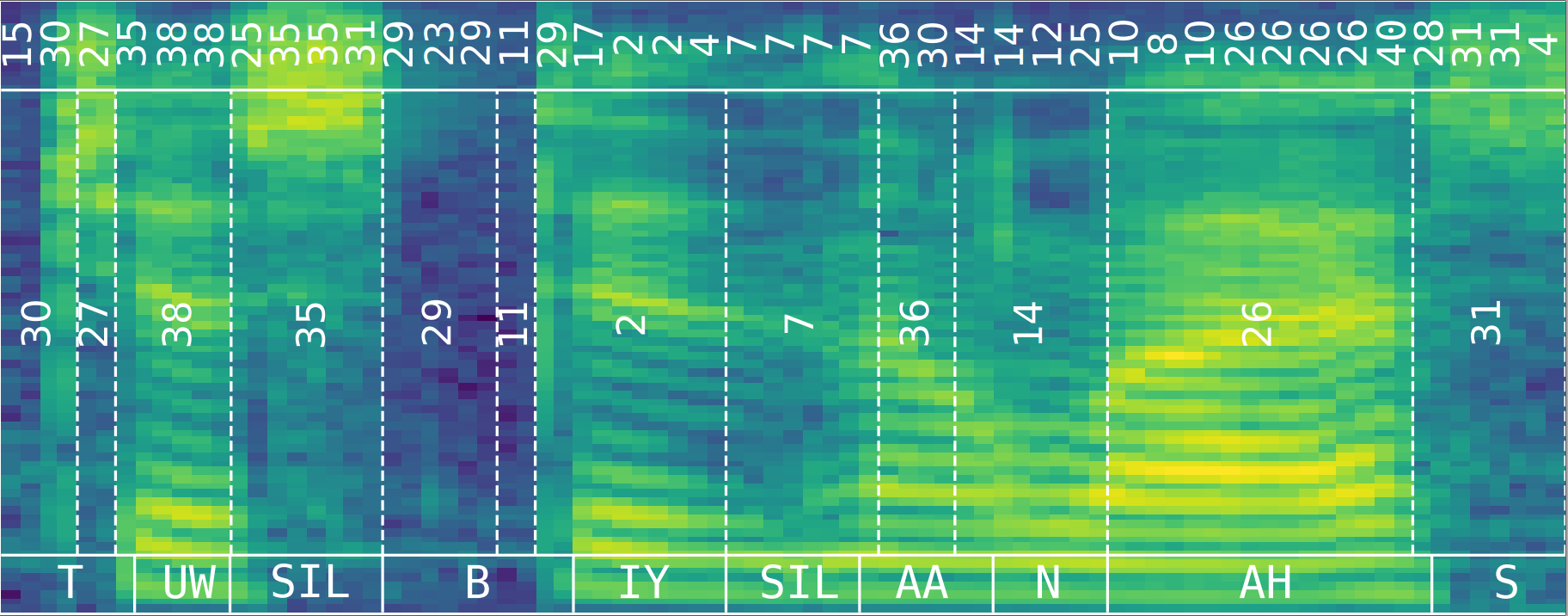}
	\caption{Spectrogram of a phrase ‘To be honest’ with the code indices (top) and the ground truth phonemes (bottom). The indices with the dashed segmentation (middle) is the result of applying our context clustering segmentation.}
	\label{fig:al}
\end{figure}

Yet, forced aligners are preciser than unsupervised ones, \textit{CL} increases performance by a significant margin, which is close to current forced aligners.
When we conducting experiments, we noticed an precision deduction when input speeches being longer. This is inevitable because of the accumulative segment differences that the monotonic clustering brings.
We also visualize an example in Fig.~\ref{fig:al}.
It's worth-noticing that our method tends to generate a more narrow segment than ground-truth. This further proves that our codebook is sufficient for capturing the main acoustic character of phonemes. When performing the robust clustering, those non-prominent frames are discarded.

\begin{table}[h]
	\caption{Phoneme boundary segmentation results on TIMIT test sets. CL denotes the CLustering method in Sec.~\ref{cl}.}	
	\centering
	\resizebox{\linewidth}{!}{
		\begin{tabular}{lccccc}
			\toprule
			Methods & Prec. & Rec. & \emph{F} & \emph{R}-val & Overlap \\ \midrule
			\rowcolor{gray!40}
			Supervised seg. &  &  &  &  &  \\ \hline
			MFA~\cite{mfa} & 0.62 & 0.63 & 0.63 & 0.68 & 75.0\% \\
			WebMAUS~\cite{web} & 0.70 & 0.70 & 0.70 & 0.75 & 78.8\% \\ \midrule
			\rowcolor{gray!40}
			Unsupervised seg. 10ms-tolerance &  &  &  &  &  \\ \hline
			IRRM & 0.28 & 0.33 & 0.27 & 0.22 & 34.7\% \\ 
			IRRM + Greedy \textit{N}-seg & 0.40 & 0.42 & 0.41 & 0.56 & 57.3\% \\ 
			IRRM + CL & 0.60 & 0.63 & 0.61 & 0.66 & 70.3\% \\ \bottomrule
	\end{tabular}}
\end{table}

\section{Conclusions}
\label{con}
This work demonstrates the feasibility of invariant representations modeling from accented speeches.
We utilize a codebook to learn the articulatory features for each phoneme in the latent space. In the wav2vec 2.0-way of pre-training, IRRM could substantially learn the phonetic units from audios, thus bridging the gap between texts and speeches.
IRRM outperforms the baseline both in-domain modeling ability and cross-domain adaptation ability.
Among future research lines, we would further improve the modeling ability of the learned representations. The main challenge is capturing phonetic characteristics invarient of all kinds of interference factors, such as noise, duration, energy of the raw speech.

\section{Acknowledgment}
This paper is supported by the Key Research and Development Program of Guangdong Province under grant No.2021B 0101400003. Corresponding author is Jianzong Wang from Ping An Technology (Shenzhen) Co., Ltd (jzwang@188.com).

\vfill
\pagebreak

\bibliographystyle{IEEEbib}
\bibliography{mybib}

\end{document}